\documentstyle[preprint,tighten,prc,aps]{revtex}
\begin{document}
\draft
\preprint{\vbox{\noindent 
 \hfill LA-UR-97-1375\\
 \null\hfill nucl-th/9704063}}
\title{Parity conserving $\bbox{\gamma}$ asymmetry in n-p 
radiative capture}
\author{Attila Cs\'ot\'o and B.\ F.\ Gibson}
\address{Theoretical Division, Los Alamos National 
Laboratory, Los Alamos, New Mexico 87545}
\author{G.\ L.\ Payne}
\address{Department of Physics \& Astronomy, University of 
Iowa, Iowa City, Iowa 52242}
\date{28 April, 1997}

\maketitle

\begin{abstract}
The importance of n-p radiative capture, utilizing 
polarized cold neutrons, as a means of measuring the weak 
pion coupling constant is reviewed.  Parity conserving 
processes of the form 
${\bf k}_{\gamma} \cdot ({\bf s}_n \times {\bf k}_n)$
can contribute to the ${\bf s}_n \cdot {\bf k}_{\gamma}$
photon asymmetry in any such experiment, if the
apparatus is not perfectly symmetric.
For an incident laboratory neutron energy of 
0.003 eV a value of $A^{PC}_{\gamma}=0.67 \times 10^{-8}$ 
is obtained for two different potential models (Argonne 
AV14 and Nijmegen Reid93). Serving as an extreme test case, 
the Reid soft core potential yields $0.61\times 10^{-8}$, 
close to the result of the contemporary forces. Implications 
for extracting the weak pion coupling constant and for 
monitoring the beam polarization are discussed.
\end{abstract}
\narrowtext

\section{Introduction}

It was more than 60 years ago that the first photonuclear 
experiment was performed \cite{Cha35}: 
$^2{\rm H} + \gamma \rightarrow {\rm n} + {\rm p}$.  Over 
the following 15 years, it came to be realized \cite{Aus50} 
that the approximately 330 mb cross section for the inverse 
reaction (thermal neutron capture by hydrogen) 
\cite{Noy65,Cox65} was some 10\% larger than that for which 
theoretical models could account. This discrepancy between 
experiment and theory produced the first incontrovertible 
evidence for the importance of meson exchange currents in 
nuclear reactions \cite{Bro72,Gar73}. Some 20 years ago it 
was observed that deuteron photodisintegration yielded forward 
going ($0^{\circ}$) protons \cite{Hug76} in greater numbers
than nonrelativistic theory predicted.  This was 
subsequently confirmed by two independent measurements 
\cite{Gil81,Mey84}. The observation was demonstrated to 
provide evidence for the contribution of the relativistic 
spin-orbit dipole operator even at photon energies well 
below 50 MeV \cite{Cam82,Fri84}. These are but two examples 
that illustrate the significant role played by low-energy 
neutron-proton radiative capture and threshold deuteron 
photodisintegration experiments in developing our 
understanding of nuclear physics in terms of the physically 
observable hadrons, the baryons and mesons, and their 
interactions.

The deuteron has also played an important role in efforts 
to understand the weak interaction in nuclei. In the same 
year that parity nonconservation (PNC) was observed in 
$\beta$ and $\mu$ decay, a first search for parity 
violation in the NN interaction was reported \cite{Tan57}.
The first evidence for such parity nonconservation was 
found in the radiative capture of neutrons by 
$^{181}$Ta \cite{Lob67}. The observed signal was of the 
expected size:
\begin{equation}
V^{PNC}_{NN}/V^{PC}_{NN} \sim G m^2_{\pi} \sim 10^{-7},
\end{equation}
where $G = 1.01 \times 10^{-5}/M^2_N$ is the weak coupling 
constant. The neutron capture technique was later extended 
in an attempt to detect the parity nonconserving circular 
polarization signal in n-p radiative capture. 
Unfortunately, the initial measurement was contaminated by 
circularly polarized photons from bremsstrahlung, but an improved 
measurement \cite{Kny84} yielded a final upper limit of 
$P_{\gamma} = (1.8 \pm 1.8) \times 10^{-7}$.  The nominal 
value was consistent with theoretical expectations of 
$0.6 \times 10^{-7}$. However, to obtain an accuracy sufficient 
to extract more than an upper limit on the weak coupling 
constants via this method would be difficult, because of the 
small analyzing power (0.045) of the $\gamma$-ray polarimeter.

A summary of efforts over the years to understand the 
coupling constants defining the weak Hamiltonian $H_W$ can 
be found in the reviews by Adelberger and Haxton \cite{Ade85} 
and by Haeberli and Holstein \cite{Hae95}. Various combinations 
of the weak meson-exchange coupling constants \cite{DDH80} 
($f_{\pi}, h^0_{\rho}, h^1_{\rho}, h^2_{\rho}, h^0_{\omega}, \ldots$) 
contribute to a number of scattering 
and reaction processes. However, most PNC measurements are 
sensitive to a linear combination of these parameters or to 
only the $\Delta I = 0,2$ components. In particular, the 
n-p radiative capture $P_{\gamma}$ measurements and 
possible helicity measurements are sensitive to 
$\Delta I = 0,2$ mixing effects in the $^1$S$_0-^3$P$_0$ and 
$^3$S$_1-^3$P$_1$ amplitudes.  Similarly, the PNC effects 
in low-energy pp scattering (at LAMPF \cite{Nag79} and 
PSI \cite{Bal84}) explored primarily $\Delta I = 0,2$ but 
failed to investigate $\pi$-exchange effects. (The 
$f_{\pi}$ coupling does not enter the picture in the case 
that projectile and target are identical.)  Alternatively, 
a measurement of the photon emission asymmetry 
$A^{PNC}_{\gamma}$ when polarized neutrons are captured by 
protons is sensitive to $\Delta I = 1$ mixing effects in 
the $^3$S$_1-^3$P$_1$ amplitude and, thus, to the weak pion 
coupling constant, $f_{\pi}$. ($P_{\gamma}$ and 
$A_{\gamma}$ are essentially independent.)  The dependence
of this photon asymmetry on the weak coupling constants is
calculated to be \cite{Hae95,Des80} 
\begin{eqnarray}
A^{PNC}_{\gamma} &=& -0.107 \, f_{\pi} + 0.003 \, 
h^1_{\omega} - 0.001 \, h^1_{\rho} \nonumber \\
&\simeq& -0.11 \, f_{\pi} + (\rm{negligible} \; \rho/\omega 
\; \rm{contributions}) \, .  
\end{eqnarray}
\noindent
A definitive measurement of the parity violating component 
of $A_{\gamma}$ would test whether $f_{\pi}$ agrees with 
the the neutral-current-enhanced weak current prediction 
of Desplanques, Donoghue, and Holstein \cite{DDH80} 
($-0.5 \times 10^{-7}$) or is, in fact, significantly 
smaller. Studies of a parity mixed doublet in $^{18}$F
indicate a strong suppression of $f_\pi$ relative to the
DDH best values \cite{Hae95,Hor95}. A comparison of 
$f_{\pi}$ extracted from $A_{\gamma}$ in the n-p radiative 
capture reaction with that coming from measurements in 
$^{18}$F and in other light and heavy nuclei would provide 
a first insight regarding possible modification of 
$f_{\pi}$ when embedded in the nuclear medium.

The measurement of $A_{\gamma}$ in n-p radiative capture 
reached feasibility with the development of intense cold, 
polarized neutron beams from the high flux reactor at ILL 
(Institut Laue-Langevin). The initial result was reported 
\cite{Cav77} to be 
$A_{\gamma} = (-6 \pm 21) \times 10^{-8}$. 
The final analysis provided only a refined upper 
limit \cite{Alb88} of 
$A_{\gamma} = (-1.5 \pm 4.7) \times 10^{-8}$. 
Because the experiment was limited by statistical 
uncertainties, it is anticipated that longer running might 
achieve an order of magnitude improvement. Such a 
measurement could answer the question of whether there 
exists a significant neutral current enhancement of 
$f_{\pi}$ in the NN weak interaction.

The experimental situation which we study in this paper is
the following. A transversely polarized neutron beam is
incident on an unpolarized proton target. The $z$ axis of
our coordinate system is defined by the neutron momentum, 
${\bf k}_n$, while the direction of the neutron 
polarization, ${\bf s}_n$, is parallel to the $x$ axis. The
momentum of the outgoing photon, ${\bf k}_\gamma$ makes an
angle $\theta$ with ${\bf k}_n$, while $\phi$ is the angle
between the plane of ${\bf k}_n$ and ${\bf k}_\gamma$ and
the plane defined by ${\bf k}_n$ and ${\bf s}_n$. All angles 
are given in the center-of-mass frame, where the total
momentum is zero; all energies and momenta are specified in 
the laboratory frame, where the proton is at rest.

The observable in which we are interested is $A_\gamma$, the
asymmetry of the photon distribution with respect to the
polarization direction.  The only parity conserving (PC) 
scalar, built from ${\bf k}_n$, ${\bf s}_n$, and 
${\bf k}_\gamma$, describing this situation, is 
${\bf s}_n\cdot[{\bf k}_n\times{\bf k}_\gamma]$, while 
the only parity nonconserving (PNC) pseudoscalar is 
${\bf s}_n\cdot{\bf k}_\gamma$.  The former leads to a
left-right ($\sin\theta\sin\phi$) asymmetry, while the latter leads 
to an up-down ($\sin\theta\cos\phi$) asymmetry. Because of these 
different symmetry properties, the two can be separated 
exactly in a perfect detector.  However, if there is a small 
asymmetry in the left-right spatial acceptance of the 
detector, then $A^{PC}_\gamma$ leads to an $A^{PNC}_\gamma$-like 
background.  The asymmetry in the up-down or left-right 
acceptance is typically less than 1\% in a carefully 
constructed detector.  The question is whether this is 
adequate to avoid any contamination of $A^{PNC}_\gamma$ 
from $A^{PC}_\gamma$.  In the ILL experiment it was 
{\it assumed} that contamination of $A^{PNC}_{\gamma}$ due 
to a parity conserving asymmetry, $A^{PC}_{\gamma}$, was 
negligible.  Whether such an assumption was at that time
warranted, any effort to push such a measurement to obtain 
more than an upper limit should take into account 
$A^{PC}_{\gamma}$. 

Our goal here is to calculate $A^{PC}_{\gamma}$, using 
contemporary nucleon-nucleon potential models.  We seek to 
provide an estimate upon which to base any new measurement 
of the parity nonconserving photon angular asymmetry in n-p 
radiative capture in order to determine a value for the 
pion weak coupling constant.  We want to understand how 
robust is the estimate of $A^{PC}_{\gamma}$ in order to shed 
light on two important questions: (1) If the measured 
asymmetry that determines $A^{PNC}_{\gamma}$ is 
sufficiently small that one must separate the PNC and PC 
asymmetries, then how large might the PC contamination be? 
(2) If the separation can be achieved experimentally, then 
can $A^{PC}_{\gamma}$ be used as a polarization monitor in 
the measurement?  An {\it in situ} polarization monitor would
considerably enhance the reliability of any measurement.

This short paper is structured as follows: In the next 
section, we outline the expressions required to calculate 
the parity conserving photon asymmetry.  In the following 
section we discuss briefly our method for obtaining the 
nucleon-nucleon bound state and continuum wave functions. 
In the last section of the manuscript we summarize our 
results, compare them with the known parity violating 
asymmetry estimates, and discuss the implications of our 
calculations for future measurements of $f_{\pi}$.

\section{Angular distribution}

For the cold neutron energies of interest, only the lowest 
multipoles in the radiative capture process are important. 
The angular distribution has the general sin$^2\theta$ 
form characteristic of an $E1$ dominated transition.  The 
angular asymmetry arises from an $E1 \times M1$ interference, 
so that the photon asymmetry exhibits an expected sin$\phi$ 
dependence.  These properties have been thoroughly discussed 
in the literature.

We adopt the nonrelativistic phenomenological treatment of 
n-p radiative capture published by Partovi \cite{Par64} (and 
use $\hbar=c=1$ units). The primary approximation is the 
neglect of any nucleon structure and meson exchange currents. 
The parity conserving n-p capture cross section can be 
expressed as
\begin{equation}
\frac{d \sigma}{d \Omega} = I_0(\theta)[1 + P_t \; 
B(\theta)\;  \rm{sin} \phi],
\label{cross}
\end{equation}
where $P_t$ is the transverse polarization of the incident 
nucleon, and $\theta$ and $\phi$ are defined above. 

The function $I_0(\theta)$ appearing which determines the
unpolarized differential cross section can be written as:
\begin{eqnarray}
I_0(\theta) &=& \frac{\omega^2}{4 k^2} 
      \sum^2_{M=0} \; \sum_{L'=L} \; \sum_{L=1} 
      (2-\delta_{M0})(2-\delta_{LL'})              {\nonumber} \\
    & & \times \{[d^{(L)}_{1M}(\theta) d^{(L')}_{1M}(\theta)
      + d^{(L)}_{1-M}(\theta) d^{(L')}_{1-M}(\theta)]  {\nonumber} \\
    & & \hspace*{.5cm} \times {\rm Re} \sum^1_{s=0} \, \sum^1_{m^d=-1} 
             [(s \; M+m^d | E^{(L)} |m^d)(s \; M+m^d | E^{(L')} |m^d)^* 
                                                 {\nonumber} \\
    & & \hspace*{2.3cm} + (s \; M+m^d | M^{(L)} |m^d)
                            (s \; M+m^d | M^{(L')} |m^d)^*] \\
    & & + [d^{(L)}_{1M}(\theta) d^{(L')}_{1M}(\theta)  
       -  d^{(L)}_{1-M}(\theta) d^{(L')}_{1-M}(\theta)] \nonumber \\
    & & \hspace*{.5cm} \times {\rm Re} \sum^1_{s=0} \, \sum^1_{m^d=-1} 
              [(s \; M+m^d | E^{(L)} |m^d)(s \; M+m^d | M^{(L')} |m^d)^*
                                                 {\nonumber} \\
    & & \hspace*{2.3cm} + (s \; M+m^d | M^{(L)} |m^d)
                            (s \; M+m^d | E^{(L')} |m^d)^*] \} \nonumber .
\end{eqnarray}
\noindent
Here $k$ is the relative n-p momentum, $d_{mm'}^{(J)}$ is 
the reduced rotation matrix, and the electric and magnetic 
multipoles are obtained by the usual expansion of 
$\bbox{\epsilon}\exp(i{\bf k}_{\gamma} \cdot \bbox{\xi)}$ 
in terms of the photon polarization 
$\bbox{\epsilon}$, the nucleon coordinate $\bbox{\xi}$, 
and the photon momentum (energy) ${\bf k}_{\gamma}$ ($\omega$):
\begin{eqnarray}
\bbox{\epsilon} \; e^{(i {\bf k}_{\gamma} \cdot \bbox{\xi})} 
   &=& \sum_{L \, M} D^{(L)}_{M \mu}(0,-\theta,-\phi) 
        \left(\frac{2 \pi (2L+1)}{L(L+1)} \right)^{1/2} \nonumber \\
   & & \times \left\{ - \frac{i^{L+1}}{\omega} \bbox{\nabla} 
       \left[ \left(1 + \xi\frac{d}{d \xi} \right) j_L(\omega \xi)
        Y^{(L)}_{M}(\theta,\phi) \right] \right. \\
   & &  \left.  - \; i^{L+1} \omega \bbox{\xi} j_L(\omega \xi) 
        Y^{(L)}_{M}(\theta,\phi)  
       \; - \; \mu i^L j_L(\omega \xi) [ \bbox{L} 
        Y^{(L)}_{M}(\theta,\phi)] \right\}. \nonumber
\end{eqnarray}
\noindent
The $D^{(L)}_{M \mu}$ function appearing above is the 
standard rotation matrix.  The first two terms in the 
summation constitute the electric multipoles; the last term 
is the magnetic multipole.  The first term gives rise to 
usual Sigert theorem electric multipole operators; the 
second term generates the electric spin-dependent operator 
as well as a retardation correction to the electric 
multipole transitions.

The term giving rise to the photon asymmetry in Eq.(5) can 
be expressed as follows:
\begin{eqnarray}
I_0(\theta) B(\theta) &=& 
    -\frac{\omega^2}{\sqrt{2} k^2} \; {\rm Im} \sum_{L \, L'' \, m^d} 
       \nonumber \\
    & & \{[d^{(L)}_{1 \, -m^d}(\theta) d^{(L')}_{1 \, 1-m^d}(\theta)
      - d^{(L)}_{1 \, m^d}(\theta) d^{(L')}_{1 \, m^d-1}(\theta)]  
                                                 \nonumber \\
    & & \hspace*{.5cm} [(10| E^{(L)} |m^d)(11| E^{(L')} |m^d)^* 
                                                 \nonumber \\
    & & \hspace*{2.3cm} + (10| M^{(L)} |m^d)(11| M^{(L')} |m^d)^*] 
                                                 \nonumber \\
    & & \hspace*{.5cm} - (00| E^{(L)} |m^d)(11| E^{(L')} |m^d)^* 
                                                 \nonumber \\
    & & \hspace*{2.3cm} - (00| M^{(L)} |m^d)(11| M^{(L')} |m^d)^*] \\
    & & + [d^{(L)}_{1 \, -m^d}(\theta) d^{(L')}_{1 \, 1-m^d}(\theta)  
       +  d^{(L)}_{1 \, m^d}(\theta) d^{(L')}_{1 \, m^d-1}(\theta)] 
                                                  \nonumber \\
    & & \hspace*{.5cm} [(10| E^{(L)} |m^d)(11| M^{(L')} |m^d)^*
                                                  \nonumber \\
    & & \hspace*{2.3cm} + (10| M^{(L)} |m^d)(11| E^{(L')} |m^d)^*] 
                                                  \nonumber \\
    & & \hspace*{.5cm} - (00| E^{(L)} |m^d)(11| M^{(L')} |m^d)^*
                                                  \nonumber \\
    & & \hspace*{2.3cm} - (10| M^{(L)} |m^d)(11| E^{(L')} |m^d)^*]
    \}. 
                          \nonumber
\end{eqnarray}
\noindent
The relative n-p momentum $k$ is related to the photon
energy by 
\begin{equation}
k^2 \; = \; M \left[ \omega \left(1 \; - \; 
\frac{E_d}{2M} \right)\; - \; E_d \left(1 \; - \; 
\frac{E_d}{4M} \right) \right], 
\end{equation}
where $E_d$ is the deuteron binding energy and $M$ is the
nucleon mass.

At very low energies, using the properties of the $d$
functions and the EM matrix elements, $B(\theta)$ in Eq.\ 
(\ref{cross}) reduces to $B(\theta)=A^{PC}_\gamma\sin\theta$, 
where $A^{PC}_\gamma$ is the parity conserving $\gamma$ asymmetry 
in which we are interested.  Thus we recover the 
${\bf s}_n\cdot{\bf k}_\gamma$ angular distribution, discussed 
above. 

\section{Wave function solution}

The outgoing scattering wave functions are evaluated by numerically 
solving the two-body Schr\"odinger equation for each partial wave. 
For each value of the total angular momentum $j$, there are four 
equations corresponding to the four possible combinations of orbital 
angular momentum $l$ and spin $s$.  Two of these equations, 
($l=j, s=0$) and ($l=j, s=1$), are uncoupled.  Following Partovi
\cite{Par64}, we label them by $\lambda=2$ and $\lambda=4$.  The 
other two equations, ($l=j-1, s=1$) and ($l=j+1, s=1$), are coupled 
and have two independent sets of regular solutions.  These are 
labeled by $\lambda=1$ and $\lambda=3$, where $\lambda=1$ is the 
solution that reduces to the $l=j-1$ function when there is no 
coupling.  The other solution $\lambda=3$ reduces to the $l=j+1$ 
function in this limit.  The boundary conditions for the reduced 
partial wave functions are such that the functions are zero at the 
origin and have the asymptotic form
\begin{equation} 
v_{ls\lambda}^j(kr) \rightarrow 
            \sin(kr-{1\over 2}l\pi+\delta_{\lambda}^j) \; . 
\end{equation}
The numerical solutions were obtained by expanding each function in a 
complete set of cubic splines \cite{Pre75} and then using the 
collocation method to generate a matrix equation for the coefficients 
of the spline expansion.  The eigenvalue coupled equations for the 
bound-state wave functions are solved in an analogous manner, except 
that the boundary condition in the asymptotic region requires that 
the functions go to zero.

The wave functions were first tested for accuracy by comparing with 
the numerical results reported by Partovi.  Agreement of better than 
1\% was obtained with his matrix elements for various partial waves.
In addition, we reproduced his published polarization functions.
Finally, we obtained excellent agreement with the total capture
cross section for thermal neutrons reported by Arenh\"ovel and 
Sanzone \cite{Are90} using two different potential models 
\cite{Rei68,Wir84}.

\section{Numerical results}

We considered the Reid Soft Core (RSC) \cite{Rei68}, the Argonne 
V$_{14}$ (AV14) \cite{Wir84}, and the Nijmegen Reid93 \cite{Nij93}  
potential models to represent nucleon-nucleon scattering and the 
deuteron bound state. The RSC model was constructed to fit p-p 
scattering data in the $^1$S$_0$ 
channel, whereas the AV14 and Reid93 models were fitted to n-p 
scattering data in that channel.  The simple form of the RSC model 
makes it a good test case for numerical checks.  Moreover, by using 
the RSC p-p based force, we also obtain the widest possible range 
of values for $A^{PC}_{\gamma}$. That is, the Coulomb corrected 
p-p scattering length is $a^C_{pp} \simeq -17$ fm in contrast to 
the n-p spin-singlet scattering length which is 
$a^s_{np} \simeq -23.7$ fm.  However, because the photon asymmetry 
is a ratio of matrix elements, the dependence on the $^1$S$_0$ 
channel interaction largely cancels.  The RSC, AV14, and Reid93 
forces yield 
$A^{PC}_\gamma=0.607\times 10^{-8}$, 
$A^{PC}_\gamma=0.668\times 10^{-8}$, and 
$A^{PC}_\gamma=0.665\times 10^{-8}$, respectively. 

Taking into account the spatial resolution of a typical detector, 
this means that $A^{PC}_\gamma$ should not significantly contaminate 
a measurement of $A^{PNC}_\gamma$ unless the latter should prove to
be an order of magnitude smaller than expected; that is, unless 
$A^{PNC}_\gamma$ is found to be of the order of $10^{-9}$.  

Likewise, the theoretical estimate of $A^{PC}_{\gamma}$ appears to 
be sufficiently model independent that measurement of this quantity 
can serve as a valid polarization monitor in experiments involving 
polarized cold neutrons, if an experimental sensitivity of a
few times 10$^{-9}$ can be obtained.

Finally, the dependence of $A^{PC}_{\gamma}$ upon the neutron 
energy was found to be linear, within the region where
$B(\theta) = A^{PC}_{\gamma}\sin\theta$ holds,
for several orders of magnitude above threshold.  Thus, one 
can easily explore the photon asymmetry at energies convenient to 
a given experiment.

\acknowledgments
The work of A.\ C.\ and B.\ F.\ G.\ was performed under 
the auspices of the U.\ S.\ Department of Energy. That of 
G.\ L.\ P.\ was supported in part by the U.\ S.\ Department 
of Energy. A.\ C.\ also acknowledges support from OTKA 
Grant F019701. We are indebted to J.\ D.\ Bowman for 
introducing us to the experimental aspects of the problem 
and to P.\ Herczeg for useful discussions on weak 
interaction physics.

\end{document}